\documentstyle[prl,aps,multicol,epsfig]{revtex}

\begin{document}
\large
\title{Use of Quadrupolar Nuclei for Quantum Information processing by Nuclear Magnetic Resonance: Implementation
 of a Quantum Algorithm.}
\author{Ranabir Das$^\dagger$ and Anil Kumar $^{\dagger,\ddagger}$\\
        $^{\dagger}$ {\it Department of Physics,}
        $^{\ddagger}$ {\it Sophisticated Instruments Facility}\\
        {\it Indian Institute of Science, Bangalore 560012 India}\\}
\maketitle
\begin{abstract}
 Physical implementation of Quantum Information Processing (QIP) by liquid-state  
Nuclear Magnetic Resonance (NMR), using weakly coupled spin-1/2 nuclei of a molecule, is well established. 
 Nuclei with spin$>$1/2 oriented in liquid crystalline matrices is another possibility. 
 Such systems have multiple qubits per nuclei and large quadrupolar couplings resulting in well separated 
lines in the spectrum. So far, creation of pseudopure states and logic gates have been demonstrated in such systems using 
transition selective radio-frequency pulses. In this paper we report two novel developments. First, 
we implement a quantum algorithm which needs coherent superposition of states. 
Second, we use evolution under quadrupolar coupling to implement multi qubit gates. 
We implement Deutsch-Jozsa algorithm on a spin-3/2 (2 qubit) system. The controlled-not operation 
needed to implement this algorithm has been implemented here by evolution under the quadrupolar 
Hamiltonian. This method has been implemented for the first time in quadrupolar systems. Since the quadrupolar coupling  
is several orders of magnitude  greater than the coupling in weakly coupled spin-1/2 nuclei, the gate time decreases, 
increasing the clock speed of the quantum computer.
\end{abstract}
\section{Introduction}
In 1985 David Deutsch suggested an algorithm which demonstrated the use of "massive quantum 
parallelism" inherent in quantum systems \cite{deu}. Better known as Deutsch-Jozsa (DJ)
algorithm, it can distinguish between a `constant' and 
a `balanced' function in an N-qubit system in one function call, where as its classical counterpart 
 requires on the average ($2^{N-1}+1$) function calls \cite{deu,deujoz}. 
 Over the years quantum information processing have been demonstrated in 
various physical systems \cite{db,ic}. Nuclear magnetic resonance (NMR) has also 
successfully demonstrated several avenues of 
quantum information processing [7-25]. Quantum algorithms like Deutsch-Jozsa algorithm, Grover's search algorithm 
and Shor's prime factorization algorithm  
 have been successfully implemented by liquid state NMR using molecules having weakly coupled spin-1/2 nuclei 
\cite{djchu,djjo,free,kd,lo,ci,jo,ap,nat}. 
In such systems each nucleus is identified as a qubit and the coupling between 
the qubits (nuclei) is mediated through covalent bonds (indirect spin-spin J-coupling).

  A growing appreciation among researcher's is the use of quadrupolar 
nuclei with spin$>$1/2 as a suitable candidate for quantum information processing \cite{solid,fun,mulf,sim,ne,mur}. 
The energy levels of a quadrupolar nucleus are equispaced in a liquid, yielding degenerate single quantum NMR transitions. 
This degeneracy is lifted in a liquid crystalline matrix yielding $2I$ well resolved transitions, allowing
 the $2I+1$ eigenstates of a half integer spin I
 nucleus to be treated as states of an N-qubit system, provided $(2I+1)=2^N$. In such cases a single quadrupolar 
nucleus acts as several qubits \cite{fun}. In such systems, while the quadrupolar splittings are 
 of the order of several KHz, the line widths are only of few Hertz. Short and precise 
transition selective pulses can be applied to such systems \cite{fun,ne}. 

 The Hamiltonian of a quadrupolar nucleus partially oriented in liquid crystalline matrix, in the presence 
of a large magnetic field $B_0$ and  having a first-order quadrupolar coupling is given by \cite{khel}
\begin{eqnarray}
\mathcal{H}=\mathcal{H} _Z+\mathcal{H}_Q &=&-\omega_0 I_z + \frac{e^2qQ}{4I(2I-1)} (3I^2_z-I^2)S \nonumber \\
   &=& - \omega_0 I_z+\Lambda (3I^2_z-I^2),
\end {eqnarray}
 where $\omega_0=\gamma B_0$ is the resonance frequency, $\gamma$ being the gyromagnetic ratio,
 $S$ is the order parameter at the site of the nucleus,  $e^2qQ$ is the quadrupolar coupling and 
$\Lambda = e^2qQS/(4I(2I-1))$ is the effective quadrupolar coupling. 
Though $e^2qQ$ is of the order of several MHz, 
 a small value for the order parameter ($S$) converts the effective quadrupolar coupling `$\Lambda$' into several kHz. 
Preparation of pseudopure states, implementation of logic gates and  half-adder/subtracter operations and quantum simulations 
have already been demonstrated in such systems \cite{fun,mulf,sim,ne,mur}. However, so far only logical 
operations which do not require
coherent superposition have been implemented in such systems. In this work, we demonstrate that such systems can
also be utilized for quantum information processing by implementing algorithms which need coherent superpositions of states 
 such as Deutsch-Jozsa algorithm. Moreover, we propose the use of evolution under quadrupolar interaction 
for implementation of such algorithms. The Hamiltonian 
of Eq.(1) has two parts-(i) Zeeman part $(\omega_0 I_z)$ and (ii) scaled quadrupolar part
$(\Lambda (3I^2_z-I^2))$. The pulse sequence $\tau/2-(\pi)-\tau/2$ focuses the Zeeman interaction but allows the 
the system to evolve under the quadrupolar interaction.
 Similar to J-coupling, quadrupolar coupling provides interaction among multiple qubits, and can be used by such sequences
 to implement multi qubit gates. Since the gate time is inversely proportional to the strength of the interaction, 
and the scaled quadrupolar coupling is three order of magnitudes greater than J-coupling, 
the gate time decreases thereby increasing the clock speed of the quantum computer. 
However, in such systems the relaxation times are also smaller by two orders.  
 Thus decoherence takes away some of the advantage of faster gate speeds. We have experimentally implemented DJ algorithm using 
quadrupolar coupling in $^{23}$Na (spin-3/2) nuclei.

\section{Deutsch-Jozsa algorithm}
 The DJ algorithm determines the type of an unknown function when it is either constant or balanced. In the simplest
case, $f(x)$ maps a single bit to a single bit. The function is called constant if $f(x)$ is
independent of $x$ and it is balanced if  $f(x)$ is zero for one value of $x$ and unity for the other
value. For N qubit system,  $f(x_1,x_2,...x_N)$ is constant if it is independent of $x_i$ and balanced if it is zero
for half the values of $x_i$ and unity for the other half. Classically it requires ($2^{N-1}+1$) 
function calls to check if $f(x_1,x_2,...x_N)$ is constant or balanced.
 However the DJ algorithm would require only a single function call \cite{deu,deujoz}. The Cleve version of DJ algorithm
 implemented by using a unitary transformation  by the propagator
$U_f$ while adding an extra qubit, is given by \cite{cleve},
\begin{eqnarray}
\vert x_1,x_2,...x_N\rangle \vert x_{N+1}\rangle \stackrel{U_f}{\longrightarrow}
\vert x_1,x_2,...x_N\rangle \vert x_{N+1}\oplus f(x_1,x_2,...x_N)\rangle
\end {eqnarray}
The four possible functions for the single-bit DJ algorithm are given in Table 1.
\\\\
\hspace*{3cm}
TABLE 1. Constant and Balanced one-qubit functions.
\\
\hspace*{5cm}
\begin{tabular}{ccccc} \hline \hline
 &   \multicolumn{2}{c}{\ \ Constant} \quad \quad &  \multicolumn{2}{c}{\ \ \ Balanced}\\ \hline \hline
x \qquad  & \quad $f_1$ &  $f_2$ \quad \quad & \quad \quad $f_3$ & $f_4$\\ \hline
0 \qquad  & \quad $0$ &  $1$ \quad \quad  & \quad \quad $0$& $1$\\ \hline
1 \qquad  & \quad $0$ &  $1$ \quad \quad  & \quad \quad $1$& $0$\\ \hline
\end{tabular}
\\

The unitary transformations corresponding to the four possible propagators $U_f$ are
\begin{eqnarray}
U_1=\pmatrix{1&0&0&0\cr 0&1&0&0\cr 0&0&1&0\cr 0&0&0&1},~~~~~~~~~~
U_2=\pmatrix{0&1&0&0\cr 1&0&0&0\cr 0&0&0&1\cr 0&0&1&0}, \nonumber \\
\nonumber \\
U_3=\pmatrix{1&0&0&0\cr 0&1&0&0\cr 0&0&0&1\cr 0&0&1&0},~~~~~~~~~~
U_4=\pmatrix{0&1&0&0\cr 1&0&0&0\cr 0&0&1&0\cr 0&0&0&1}.
\end{eqnarray}
In the case of 2 qubits there are 2 constant and 6 balanced functions.
For higher qubits the functions are easy to evaluate using Eq.[2].

\section{DJ algorithm in spin-3/2 system}
 The Cleve version of the algorithm implemented here requires two qubits, an input qubit and a work qubit \cite{cleve}.
 Previous workers have demonstrated the algorithm using two weakly coupled spin-1/2 nuclei \cite{djchu,djjo,kd}. Here we implement
 it on a spin-3/2 nucleus.  The energy level diagram of a spin-3/2 nucleus corresponding to the Hamiltonian of Eq. [4] 
is shown in Fig. 1(a). The four 
energy levels are labeled as levels of a two qubit system (Fig. 1(a)). There 
are three single quantum transitions among the four levels, labeled as 
$\omega_{00-01}$, $\omega_{01-11}$ and $\omega_{11-10}$, Fig. 1(b) (the subscript denotes the energy levels  
between whom the transition takes place). While the above three transitions are single quantum transitions $(\Delta m=\pm 1)$ 
they are also single qubit flip transitions. However, according to the present labeling scheme 
the single qubit flip corresponding to  
$\vert 00\rangle \leftrightarrow \vert 10\rangle$ is a forbidden $ \Delta m=\pm 3$ transition. The above 
labeling scheme was chosen to optimize the experimental implementation of the algorithm. As demonstrated in 
I=7/2 systems elsewhere, optimum labelling schemes can be chosen for given logical operations \cite{mur}.  

 The quantum circuit for implementation of DJ algorithm followed in this work (Fig. 2), is similar to those used by 
 previous workers \cite{djchu,djjo}.
 The first qubit is the input qubit whereas second qubit is the work qubit. 
 The algorithm starts from a  pure state $\vert \psi\rangle=\vert 00\rangle$ followed by a Hadamard transform \cite{djjo}.
 A pseudo-Hadamard operation can be implemented by using a high power, low duration `hard' $(\pi/2)$ 
pulse along (-y) axis which creates superposition of all the qubits \cite{djchu,djjo}. In the spin-3/2
system the operator of hard $(\pi/2)_{-y}$ pulse is of the form;
\begin{eqnarray}
exp(iI_y\pi/2)=\frac{1}{2\sqrt{2}}\pmatrix{1&\sqrt{3}&\sqrt{3}&1\cr -\sqrt{3}&-1&1&\sqrt{3}\cr 
\sqrt{3}&-1&-1&\sqrt{3}\cr -1&\sqrt{3}&-\sqrt{3}&1},
\end{eqnarray}
 where the y-component of spin angular momentum in spin-3/2 system is 
\begin{eqnarray}
I_y=i \pmatrix{0&-\sqrt{3}/2&0&0\cr \sqrt{3}/2&0&-1&0\cr 0&1&0&-\sqrt{3}/2\cr 0&0&\sqrt{3}/2&0}
\end{eqnarray}
 The state of the system after $(\pi/2)_{-y}$ pulse is $\vert \psi'\rangle =exp(iI_y\pi/2)\vert \psi\rangle=
\frac{1}{2\sqrt{2}}[\vert 00\rangle -\sqrt{3}\vert 01\rangle +\sqrt{3}\vert 11\rangle -\vert 10\rangle]
=\frac{1}{2\sqrt{2}}[(\vert 0\rangle -\vert 1\rangle)(\vert 0\rangle -\sqrt{3}\vert 1\rangle)]$.  

 It is to be noted that unlike  weakly coupled spin-1/2 nuclei, 
the operator of $(\pi/2)_{-y}$ pulse used here does not create uniform superposition. 
However, it does create a coherent superposition of all the states which can be utilized for 
`quantum parallelism' as desired by the algorithm. This also works for higher spin systems like 
 spin-7/2 nuclei which can act as 3-qubit system \cite{nmrs}. After creation of  $\vert \psi'\rangle$ we 
apply the unitary operator $U_f$ which yields $\vert \psi_f''\rangle$=$U_f \vert \psi'\rangle$.  
 The operator $U_1$ is unity operator, yielding $\vert \psi_1''\rangle$=$U_1\vert \psi'\rangle$=
$\frac{1}{2\sqrt{2}}[(\vert 0\rangle -\vert 1\rangle)(\vert 0\rangle -\sqrt{3}\vert 1\rangle)]$. The operator $U_2$ flips the 
 state of the second qubit, yielding; $\vert \psi''_2\rangle=U_2 \vert \psi'\rangle=
\frac{1}{2\sqrt{2}}[(\vert 0\rangle -\vert 1\rangle)(-\sqrt{3}\vert 0\rangle +\vert 1\rangle)]$.  
  $U_3$ flips the state of second qubit only when the state of first qubit is $\vert 1\rangle$ and 
$U_4$ flips the state of second qubit only when the state of first qubit is $\vert 0\rangle$ (Table 2). 
The operators $U_3$ and $U_4$ are thus  controlled-NOT gates.  It may be noted that similar to spin-1/2 case, 
the information about the function is encoded in the relative phase of the two states of the 
input qubit; (-1) for constant and (+1) for balanced functions (Table 2). \\\\ 
TABLE 2. The function $(f)$, the operator $U_f$, and the wave function $\vert \psi''_f\rangle$ for one qubit DJ.
\hspace*{2cm}
\begin{tabular}{|ccc|} \hline
Function $(f)$ \qquad & \qquad Operator $(U_f)$ \qquad &  \qquad Wave function $\vert \psi_f''\rangle$\\ \hline
$f_1$ \qquad & \qquad $U_1$  \qquad &  \qquad  $\frac{1}{2\sqrt{2}}[(\vert 0\rangle -\vert 1\rangle)(\vert 0\rangle -\sqrt{3}\vert 1\rangle)]$ \\ 
$f_2$ \qquad & \qquad $U_2$  \qquad &  \qquad  $\frac{1}{2\sqrt{2}}[(\vert 0\rangle -\vert 1\rangle)(-\sqrt{3}\vert 0\rangle +\vert 1\rangle)]$ \\
$f_3$ \qquad & \qquad $U_3$  \qquad &  \qquad  $\frac{1}{2\sqrt{2}}[(\vert 0\rangle+\sqrt{3}\vert 1\rangle)\vert 0\rangle-(\sqrt{3}\vert 0\rangle +\vert 1\rangle)\vert 1\rangle]$ \\
$f_4$ \qquad & \qquad $U_4$  \qquad &  \qquad  $\frac{1}{2\sqrt{2}}[-(\sqrt{3}\vert 0\rangle+\vert 1\rangle)\vert 0\rangle +(\vert 0\rangle +\sqrt{3}\vert 1\rangle)\vert 1\rangle]$\\ \hline
\end{tabular}\\\\

 Density  matrices of the system confirms the different functions.
The density matrices corresponding to the states $\vert \psi_f''\rangle$ are $\sigma''_f$, given by
\begin{eqnarray}
{\raise 18mm \hbox{\hspace*{28mm}$\vert 00\rangle ~~~~~\vert 01\rangle
~~~~~\vert 11\rangle ~~~~~\vert 10\rangle$}}\hspace*{-64mm}
\sigma''_1 = \pmatrix{1&-\sqrt{3}&\sqrt{3}&-1 \cr -\sqrt{3}&3&-3&\sqrt{3} \cr 
\sqrt{3}&-3&3&-\sqrt{3} \cr -1&\sqrt{3}&-\sqrt{3}&1} 
\matrix{\vert 00\rangle \cr \vert 01\rangle \cr \vert 11\rangle \cr \vert 10\rangle}
~~~~~~~
 \sigma''_2 = \pmatrix{3&-\sqrt{3}&\sqrt{3}&-3 \cr -\sqrt{3}&1&-1&\sqrt{3} \cr
\sqrt{3}&-1&1&-\sqrt{3} \cr -3&\sqrt{3}&-\sqrt{3}&3} 
 \nonumber 
\end{eqnarray}
\begin{eqnarray} 
\sigma''_3 = \pmatrix{3&-\sqrt{3}&-3&\sqrt{3} \cr -\sqrt{3}&1&\sqrt{3}&-1 \cr
-3&\sqrt{3}&3&-\sqrt{3} \cr \sqrt{3}&-1&-\sqrt{3}&1}
 ~~~~
\sigma''_4 = \pmatrix{1&-\sqrt{3}&-1&\sqrt{3} \cr -\sqrt{3}&3&\sqrt{3}&\-3 \cr
-1&\sqrt{3}&1&-\sqrt{3} \cr \sqrt{3}&-3&-\sqrt{3}&3}
\end{eqnarray} 
 The signs of input qubit coherences $\vert 11\rangle \leftrightarrow \vert 01\rangle$ 
 and $\vert 10\rangle \leftrightarrow \vert 00\rangle$, are negative 
for $\sigma''_1$ and $\sigma''_2$ but positive for $\sigma''_3$ and $\sigma''_4$, 
indicating respectively constant and balanced functions.
   
    After $U_f$, one needs to make a measurement (Fig. 2). Theoretically this step needs a Hadamard gate followed 
by a readout of input qubit. In NMR, the Hadamard is replaced by a pseudo Hadamard, which can be implemented by a 
($\pi/2$) pulse. Similarly, the readout is also another ($\pi/2$) pulse. These two pulses cancel each other and hence in NMR, 
the result of DJ algorithm is directly available after implementation of $U_f$ \cite{djjo}. 
 As seen from Eq.[6], in the  signal acquired immediately following $U_f$; the resonance of 
 input qubit at $\omega_{01-11}$ (the central transition of Fig.1) will be of the same sign as the 
resonances of work qubit at $\omega_{00-01}$ and $\omega_{11-10}$ (the outer transitions of Fig.1) 
for constant functions, and of opposite sign for the balanced functions. It may be mentioned that only one of the 
transitions of the input qubit namely $\vert 01\rangle \leftrightarrow \vert 11\rangle$ is observed here, the 
other transition $\vert 00\rangle \leftrightarrow \vert 10\rangle$ being $\Delta m=\pm 3$.

\section{Experiment}
 The DJ algorithm is experimentally implemented here on $^{23}$Na (spin-3/2) nuclei of 
 a lyotropic liquid crystal composed of $37.9\%$ sodium dodecyl sulfate, 
$6.7\%$ decanol, and $55.4\%$ water. The liquid crystal had a nematic phase at 299K \cite{fun,sample}. 
All experiments were performed on a DRX 500MHz spectrometer.  
Fig. 1(b) shows the equilibrium spectrum consisting of three lines, with a effective quadrupolar coupling 
($\Lambda$) of about 16 kHz and integrated intensity ratio 3:4:3.

 The $\vert 00\rangle$ pseudopure state is created by applying a selective population inversion 
$(\pi)$ pulse on the $\vert 10\rangle  \leftrightarrow \vert 11\rangle$ transition followed by a 
population equilibration $(\pi/2)$ pulse on  $\vert 01\rangle  \leftrightarrow \vert 11\rangle$ 
transition and a gradient pulse to kill created coherences \cite{ernst,ne}.
  Transition selective pulses are long duration, low power r.f pulses applied at the resonant frequency between two 
energy levels, which excite a selected transition of the spectrum and leave the others unperturbed. 
Let us consider a two level sub-system $\vert i\rangle$ and $\vert j\rangle$, whose equilibrium deviation populations 
are $p_i$ and $p_j$   
\begin{eqnarray}
\sigma=\pmatrix{p_i & 0 \cr 0 & p_j}
\end{eqnarray} 
 The operator of a transition selective $(\theta)$ pulse about y-axis between these two levels would be 
\begin{eqnarray}
exp(-i I^{\vert i\rangle \leftrightarrow \vert j\rangle}_y \theta)=\pmatrix{cos(\theta/2) & sin(\theta/2) \cr -sin(\theta/2) & cos(\theta/2)}
\end{eqnarray} 
where the angular momentum operator 
 $I^{\vert i\rangle \leftrightarrow \vert j\rangle}_y =\pmatrix{0 & -i \cr i & 0}$.
 A population inversion $(\pi)$ pulse will interchange the populations between the two levels,
\begin{eqnarray}
exp(-i I^{\vert i\rangle \leftrightarrow \vert j\rangle}_y \pi).\sigma. 
exp(i I^{\vert i\rangle \leftrightarrow \vert j\rangle}_y \pi)= 
\pmatrix{0 & 1 \cr -1 & 0}.\pmatrix{p_i & 0 \cr 0 & p_j}.\pmatrix{0 & -1 \cr 1 & 0}=\pmatrix{p_j & 0 \cr 0 & p_i}. 
\end{eqnarray}
  A population equilibration $(\pi/2)$ pulse will equilibrate the populations and create coherences of the form;
\begin{eqnarray}
exp(-i I^{\vert i\rangle \leftrightarrow \vert j\rangle}_y \pi/2).\sigma.
exp(i I^{\vert i\rangle \leftrightarrow \vert j\rangle}_y \pi/2)&=& 
\frac{1}{\sqrt{2}}\pmatrix{1 & 1 \cr -1 & 1}.\pmatrix{p_i & 0 \cr 0 & p_j}.
\frac{1}{\sqrt{2}}\pmatrix{1 & -1 \cr 1 & 1} \nonumber \\ \nonumber \\
&=&\pmatrix{(p_i+p_j)/2 & (p_j-p_i)/2 \cr (p_j-p_i)/2 & (p_i+p_j)/2},
\end{eqnarray}

 which followed by gradient will retain the populations but destroy the coherences.

After the creation of pseudopure state the coherent superposition of both the qubits are  created by a non-selective 
$(\pi/2)_{-y}$ pulse. At this stage one can apply the various $U_f$. The function $U_1$ needs no pulse and the
 result given in Fig. 3 indicates that all the three transitions are of same sign and hence it is a constant function. 
 The operator $U_2$ in Eq. [3] requires two transition selective pulses $(\pi/\sqrt{3})^{\vert 00\rangle \leftrightarrow 
\vert 01\rangle}_x
(\pi/\sqrt{3})^{\vert 10\rangle \leftrightarrow \vert 11\rangle}_x$, where, 
\begin{eqnarray}
(\pi/\sqrt{3})^{\vert 00\rangle \leftrightarrow \vert 01\rangle}_x
(\pi/\sqrt{3})^{\vert 10\rangle \leftrightarrow \vert 11\rangle}_x=
\pmatrix{0&i&0&0 \cr i&0&0&0 \cr 0&0&0&i \cr 0&0&i&0}=i\pmatrix{0&1&0&0 \cr 1&0&0&0 \cr 0&0&0&1 \cr 0&0&1&0}.
\end{eqnarray}
Here we have used the fact that 
\begin{eqnarray}
(\theta)^{\vert 00\rangle \leftrightarrow \vert 01\rangle}_x= 
exp(iI_x^{00\leftrightarrow 01}\theta)=\pmatrix{cos(\sqrt{3}\theta/2)&isin(\sqrt{3}\theta/2)&0&0 \cr 
isin(\sqrt{3}\theta/2)&cos(\sqrt{3}\theta/2)&0&0 \cr 0&0&1&0 \cr 0&0&0&1}  \nonumber \\ \nonumber \\ 
(\theta)^{\vert 10\rangle \leftrightarrow \vert 11\rangle}_x=
exp(iI_x^{10\leftrightarrow 11}\theta)=\pmatrix{1&0&0&0 \cr 0&1&0&0 \cr
0&0&cos(\sqrt{3}\theta/2)&isin(\sqrt{3}\theta/2) \cr
0&0&isin(\sqrt{3}\theta/2)&cos(\sqrt{3}\theta/2)}.
\end{eqnarray}
  The result given in Fig. 3 confirms that $f_2$ is also a constant function.
  While implementing $U_3$ of Eq. [3] we note that the Pound-Overhauser CNOT gate \cite{dg} is 
similar to $U_3$, but differs from its exact form by a controlled phase operator 
\begin{eqnarray}
U_3=\pmatrix{1&0&0&0\cr 0&1&0&0\cr 0&0&0&1\cr 0&0&-1&0}\times
\pmatrix{1&0&0&0 \cr 0&1&0&0 \cr 0&0&e^{i\pi}&0 \cr 0&0&0&1}
\end{eqnarray}
 The Pound-Overhauser CNOT gate is implemented by a transition selective  
$(\pi/\sqrt{3})^{\vert 10\rangle \leftrightarrow \vert 11\rangle}_{-y}$ pulse.
 The controlled phase shift operator of Eq. [13] can be realized by using (i) only transition selective pulses 
 or (ii) transition selective pulses along with a evolution under quadrupolar coupling.

 {\bf (i) Transition selective pulse method:} Transition selective z-pulses can be used to introduce specific phases 
 to the different states \cite{ron}. For example, the  $(\phi)_z^{01\leftrightarrow 11}$
introduces a phase shift of $2\phi$ between the states $\vert 01\rangle$ and $\vert 11\rangle$.
$(\phi)_z^{01\leftrightarrow 11}$ is
implemented using three selective pulses $(\pi/4)_y (\phi)_x (\pi/4)_{-y}$ on the
transition $\vert 01\rangle \leftrightarrow \vert 11\rangle$;
\begin{eqnarray}
(\phi)_z^{01\leftrightarrow 11}=
exp(-i I_y^{01\leftrightarrow 11}\pi/4).exp(-i I_x^{01\leftrightarrow 11} \phi)
.exp(i I_y^{01\leftrightarrow 11}\pi/4)= \pmatrix{1&0&0&0 \cr 0&e^{-i\phi}&0&0 \cr 0&0&e^{i\phi}&0 \cr 0&0&0&1},
\end{eqnarray}
 where the x and y-component of the operator of transition between the states
$\vert 01\rangle \leftrightarrow \vert 11\rangle$ are
\begin{eqnarray}
I_x^{01\leftrightarrow 11}=\pmatrix{0&0&0&0 \cr 0&0&1&0 \cr 0&1&0&0 \cr 0&0&0&0} ~~~~~~~~
I_y^{01\leftrightarrow 11}=\pmatrix{0&0&0&0 \cr 0&0&-i&0 \cr 0&i&0&0 \cr 0&0&0&0}.
\end{eqnarray}
Hence the controlled phase shift operator of Eq.[13] can be achieved by using a cascade of three transition selective z-pulses 
\begin{eqnarray}
&&(\pi/4)^{\vert 00\rangle \leftrightarrow \vert 01\rangle}_{z}
(\pi/4)^{\vert 10\rangle \leftrightarrow \vert 11\rangle}_{z}(\pi/2)^{\vert 01\rangle \leftrightarrow \vert 11\rangle}_{z}
\nonumber \\ \nonumber \\
&=&\pmatrix{e^{-i\pi/4}&0&0&0 \cr  0&e^{i\pi/4}&0&0 \cr  0&0&1&0 \cr 0&0&0&1} \times
\pmatrix{1&0&0&0 \cr  0&1&0&0 \cr  0&0&e^{i\pi/4}&0 \cr 0&0&0&e^{-i\pi/4}} \times 
\pmatrix{1&0&0&0 \cr  0&e^{-i\pi/2}&0&0 \cr  0&0&e^{i\pi/2}&0 \cr 0&0&0&1} \nonumber \\ \nonumber \\
&=&e^{-i\pi/4}\pmatrix{1&0&0&0 \cr  0&1&0&0 \cr  0&0&e^{i\pi}&0 \cr 0&0&0&1}
\end{eqnarray} 
{\bf (ii)  Evolution under quadrupolar coupling  method:} 
Similar to J-coupling, evolution under the quadrupolar coupling rotates the system
about the z-axis, introducing specific phases to the states.
The quadrupolar Hamiltonian in spin-3/2 system is of the form
\begin{eqnarray}
{\mathcal H}_Q = \Lambda(3Iz^2-I^2)=3\Lambda \pmatrix{1&0&0&0 \cr 0&-1&0&0 \cr 0&0&-1&0 \cr 0&0&0&1}
\end{eqnarray}
\\
 The operator corresponding to evolution under the quadrupolar Hamiltonian for a time $\tau$
is \\
\begin{eqnarray}
e^{i{\mathcal H}_Q \tau}=exp(-i\Lambda(3Iz^2-I^2) \tau)= 
\pmatrix{e^{-i3\Lambda \tau}&0&0&0 \cr 0&e^{i3\Lambda \tau}&0&0 \cr 0&0&e^{i3\Lambda \tau}&0 \cr 0&0&0&e^{-i3\Lambda \tau}}
\end{eqnarray}
\\
 Hence the controlled phase shift operator of Eq.[13] can be achieved by
a combination $(e^{i{\mathcal H}_Q\tau})(\pi/2)^{01\leftrightarrow 11}_z$, where $\tau=\pi/12\Lambda$ is the time period of
evolution under the quadrupolar Hamiltonian.

 We have implemented  approach (ii) in our experiments. This is because, 
 the use of evolution under quadrupolar coupling reduces the number of transition selective pulses,
 enabling fast computation and less  errors due to relaxation. All the experiments were carried out 
with the carrier frequency of the r.f. pulses matching with the central transition  (on-resonance). 
 In this situation, the evolution in the rotating frame takes place only under quadrupolar Hamiltonian and the 
Zeeman term does not evolve. $U_3$ was implemented by a pulse sequence 
\begin{eqnarray}
&&(\pi/\sqrt{3})^{\vert 10\rangle \leftrightarrow \vert 11\rangle}_{-y}
(e^{i\mathcal{H}_Q\tau}) (\pi/2)^{\vert 01\rangle \leftrightarrow \vert 11\rangle}_{z} \nonumber \\ \nonumber \\
&=&\pmatrix{1&0&0&0\cr 0&1&0&0\cr 0&0&0&1\cr 0&0&-1&0}.
\pmatrix{e^{-i\pi/4}&0&0&0 \cr 0&e^{i\pi/4}&0&0 \cr 0&0&e^{i\pi/4}&0 \cr 0&0&0&e^{-i\pi/4}}. 
\pmatrix{1&0&0&0\cr 0&e^{-i\pi/2}&0&0\cr 0&0&e^{i\pi/2}&0\cr 0&0&0&1}
\nonumber \\ \nonumber\\
&=& \pmatrix{1&0&0&0\cr 0&1&0&0\cr 0&0&0&1\cr 0&0&-1&0}.
e^{-i\pi/4}\pmatrix{1&0&0&0\cr 0&1&0&0\cr 0&0&e^{i\pi}&0\cr 0&0&0&1}
=e^{-i\pi/4}\pmatrix{1&0&0&0\cr 0&1&0&0\cr 0&0&0&1\cr 0&0&1&0},
\end{eqnarray}
\\  
 where $\tau=\pi/12\Lambda$. The operator $U_4$ was implemented by a similar pulse sequence $(\pi/\sqrt{3})^{\vert 00\rangle \leftrightarrow \vert 01\rangle}_{-y}(e^{i\mathcal{H}_Q\tau}) (\pi/2)^{\vert 00\rangle \leftrightarrow \vert 01\rangle}_{-z}$, with the same  value of $\tau$. 
The result after applying $U_3$ and $U_4$ are given in Fig. 3, in which it is seen that the sign of central transition is opposite to that of the outer transitions, indicating that $f_3$ and $f_4$ are balanced functions (Eq. [6]).

  The selective excitation in this paper is achieved using Gaussian soft pulses \cite{gauss} of length 123$\mu$s. During the 
selective pulses the unexcited transitions continue to experience 
quadrupolar interaction, resulting in a rotation around the z-axis, which leads to phase
errors. To minimize such errors, the length of the selective pulses were so chosen that the 
 phase rotation is in multiples of $2\pi$ \cite{djjo}. However, errors due to relaxation could not be avoided.  
For example, the peak intensities are slightly different from the expected. 
It is because the relaxation times are: $T_1$=16ms for all the three transitions; 
$T_2$ was 14ms for the central transition and 4ms for the outer transitions.  
Since, the relaxation time of the outer transitions is less than the central, the coherences of the outer 
transitions decay faster, decreasing their  peak intensities.

  In quantum information processing by NMR, the gate time is of the order of the inverse of coupling 
  and the coherence time is proportional to inverse of the linewidth. In liquids the coupling values 
are $\sim$ 100 Hz and the linewidth $\sim$ 1 Hz yielding a gate time $\sim$ 3 ms, a coherence time $\sim$ 300 ms
 and hence a dynamic range of two oreders of magnitude. 
In the quadrupolar system described in this paper, the quadrupolar coupling value is $\sim$ 16 kHz 
yielding a  quadrupolar evolution gate time $\sim$ 20$\mu$s. Since the coherence times are 14ms and 4ms for inner and outer transitions 
respectively, the system yields a dynamic range of 3 orders of magnitude. Thus, both the dynamic range and the 
clock speed (inverse of gate time) are better in the quadrupolar system described here.
    
\section{conclusions}
 The implementation of quantum algorithms on  quadrupolar nuclei validate their use as a  alternate 
candidates
 for quantum information processing. DJ algorithm has been implemented here in a spin-3/2 system 
 by manipulation of coherent superposition using evolution under quadrupolar interaction and r.f. pulses. 
The errors in our experiments were mainly caused by relaxation and imperfection of 
r.f. pulses. Use of tailored multi frequency pulses \cite{mulf} can further decrease gate time and 
 relaxation errors. Some quantum algorithms like Grover's search algorithm require uniform superposition of states, 
which can be realized by application of  multiple quantum pulses \cite{vega}. 
 Efforts are ongoing in our lab to develop such pulses and implement various quantum algorithms in 3/2 and 7/2 spin systems.
 Since completion of this work, an implementation of continuous version of Grover's search 
algorithm has also been reported in a spin-3/2 system \cite{cong}.
`
\section{Acknowledgments}
 The authors thank T.S. Mahesh, Neeraj Sinha, N. Suryaprakash and K.V. Ramanathan 
for useful discussions. The use of DRX-500 NMR spectrometer funded by the Department of
Science and Technology, New Delhi, at the Sophisticated
Instuments Facility, Indian Institute of Science, Bangalore, is also gratefully acknowledged.

\newpage
\hspace{6cm}\large{FIGURE CAPTIONS}
\\\\
FIG. 1: (a) Energy level diagram of a spin-3/2 nucleus oriented in a liquid crystal matrix. The different
spin states can be labeled as states of a 2-qubit system.
The equilibrium deviation populations of different states under high-field high-temperature approximation 
are schematically shown by the dots on the right hand side. (b) The equilibrium spectrum of $^{23}$Na 
obtained after a hard ($\pi/2)_{y}$ pulse. Along x-axis are the frequencies in KHz and along y-axis are the intensities. The three single quantum transitions are well separated by an
effective quadrupolar coupling ($\Lambda$) of about 16 kHz. The outer lines are broader than the inner line
 (line widths are different due to differences in relaxation matrix elements as well as due to fluctuations 
in $S$ values). These fluctuations in $S$ values affect only the outer transitions in the first order, reducing 
their transverse relaxation time $T_2$. The integrated intensities are in the correct ratio of 3:4:3. 
The spectrum is plotted with a Lorentzian line-broadening factor of 200Hz.
\\\\

FIG. 2: Quantum circuit for implementing Deutsch-Jozsa algorithm. The first $(\pi/2)_{-y}$ hard pulse creates
superposition of all states. $U_f$ is the unitary transform corresponding to the function $f$. The last step
is measurement. In NMR this step can be simplified to acquisition of signal immediately after $U_f$ is implemented.
The sign of the  input qubit's resonances with respect to those of the work qubits resonances (Eq. [6]) 
distinguish between the constant and balanced functions.
\\\\
FIG. 3: Implementation of DJ algorithm on $^{23}$Na (spin-3/2) nuclei. The algorithm starts from $\vert 00\rangle$ pseudopure state. After
running through the quantum circuit of Fig.1, the acquired signal is Fourier transformed. The spectra
corresponding to the operations $U_1$, $U_2$, $U_3$ and $U_4$ are given. Along x-axis are the frequencies in KHz and along y-axis are the intensities. Constant functions are
distinguished from balanced functions by the sign of resonance of input qubit.
 As seen by the single quantum coherences of Eq. [6]; for $U_1$ and $U_2$ the resonance of input
qubit (central transition) has same sign as the resonances of work qubit (outer transitions), 
implying that the corresponding functions
$f_1$ and $f_2$ are constant, whereas for $U_3$ and $U_4$, the sign of the central transition are opposite,
indicating that $f_3$ and $f_4$ are balanced. 

\newpage
\begin{figure}
\epsfig{file=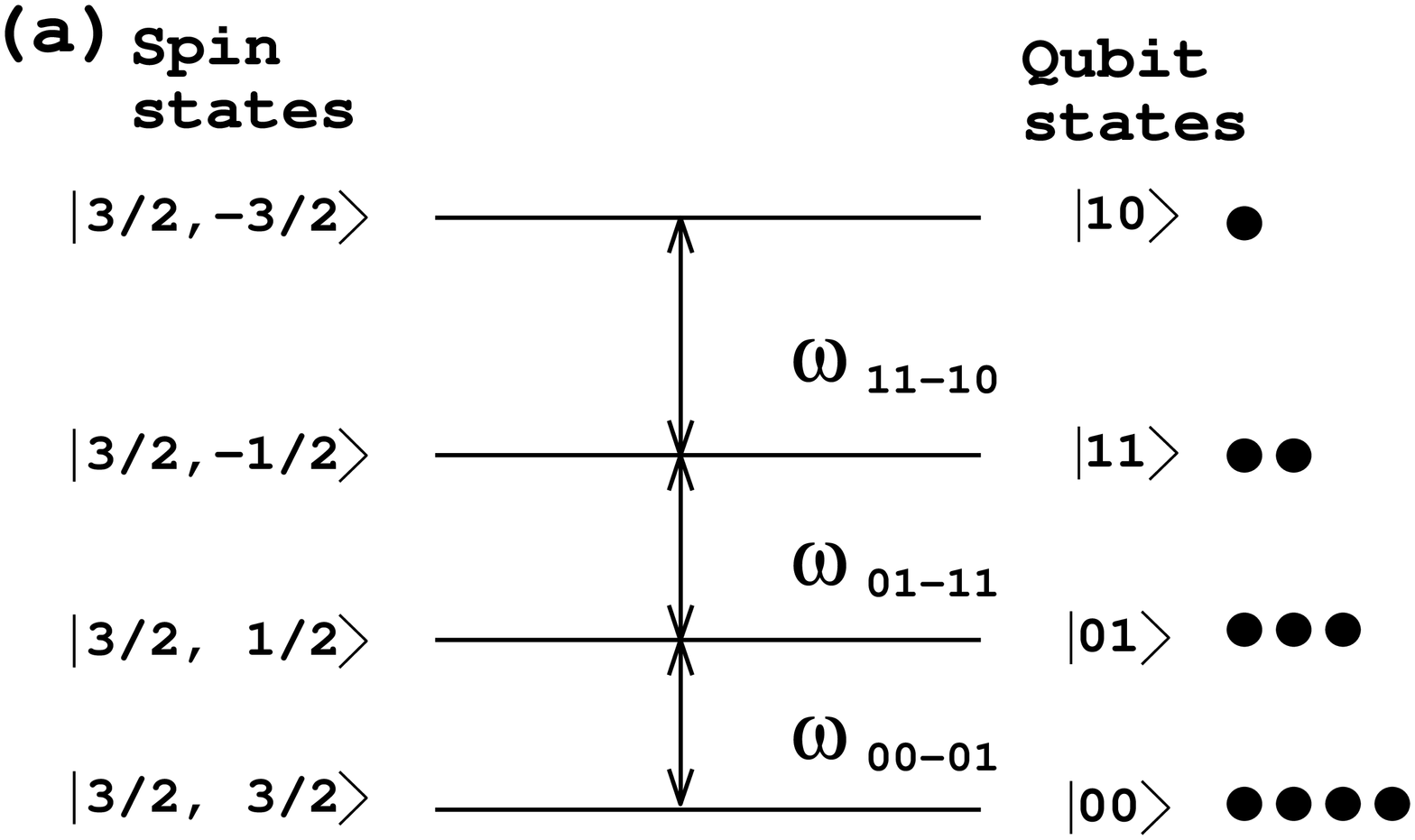,height=9cm,angle=270}
\end{figure}

\vspace*{-6cm}
\begin{figure}
\hspace{9.3cm}
\epsfig{file=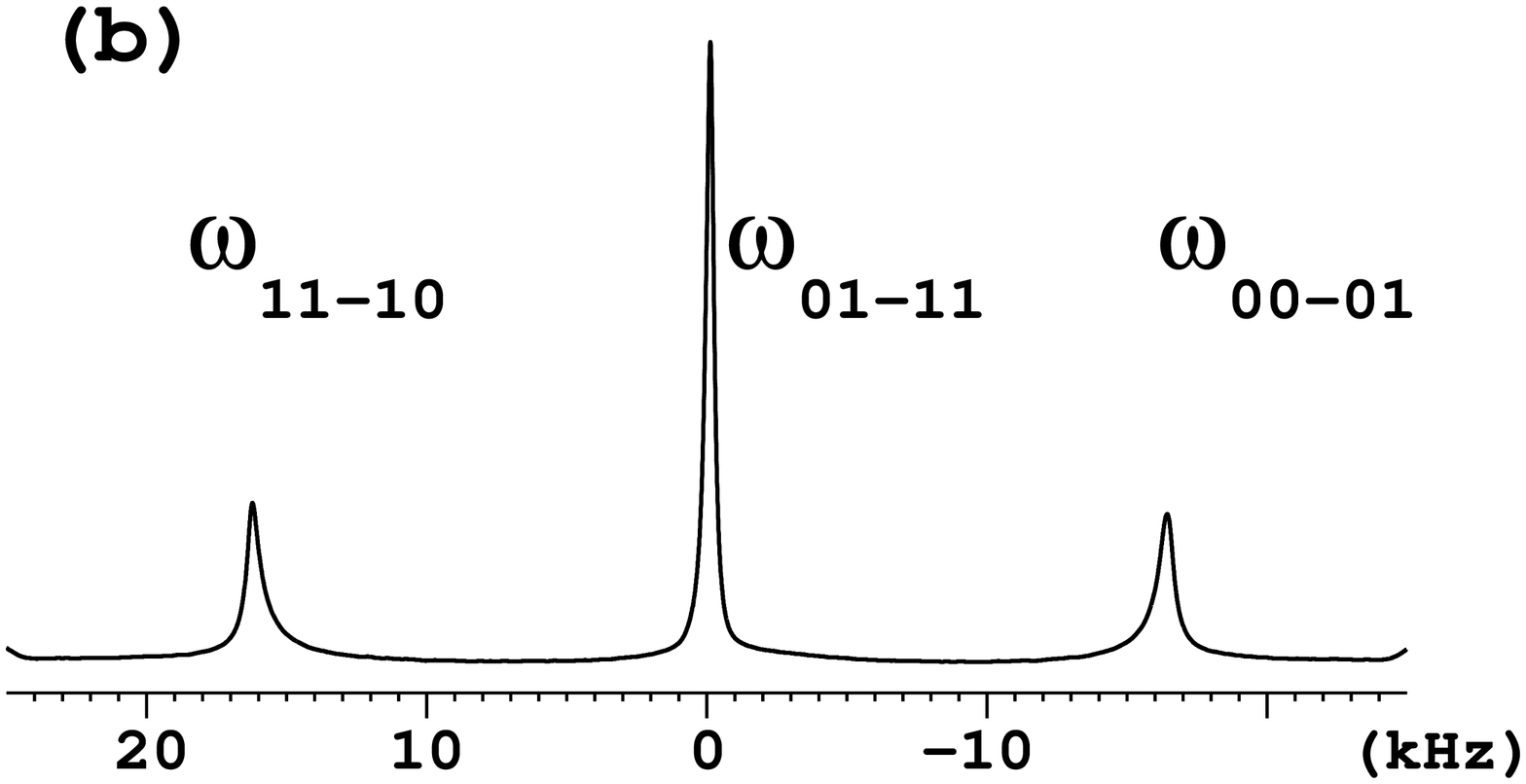,height=8cm,angle=270}
\end{figure}

\vspace{2cm}
\hspace{7cm}
\huge{Figure 1}
\newpage
\begin{figure}
\epsfig{file=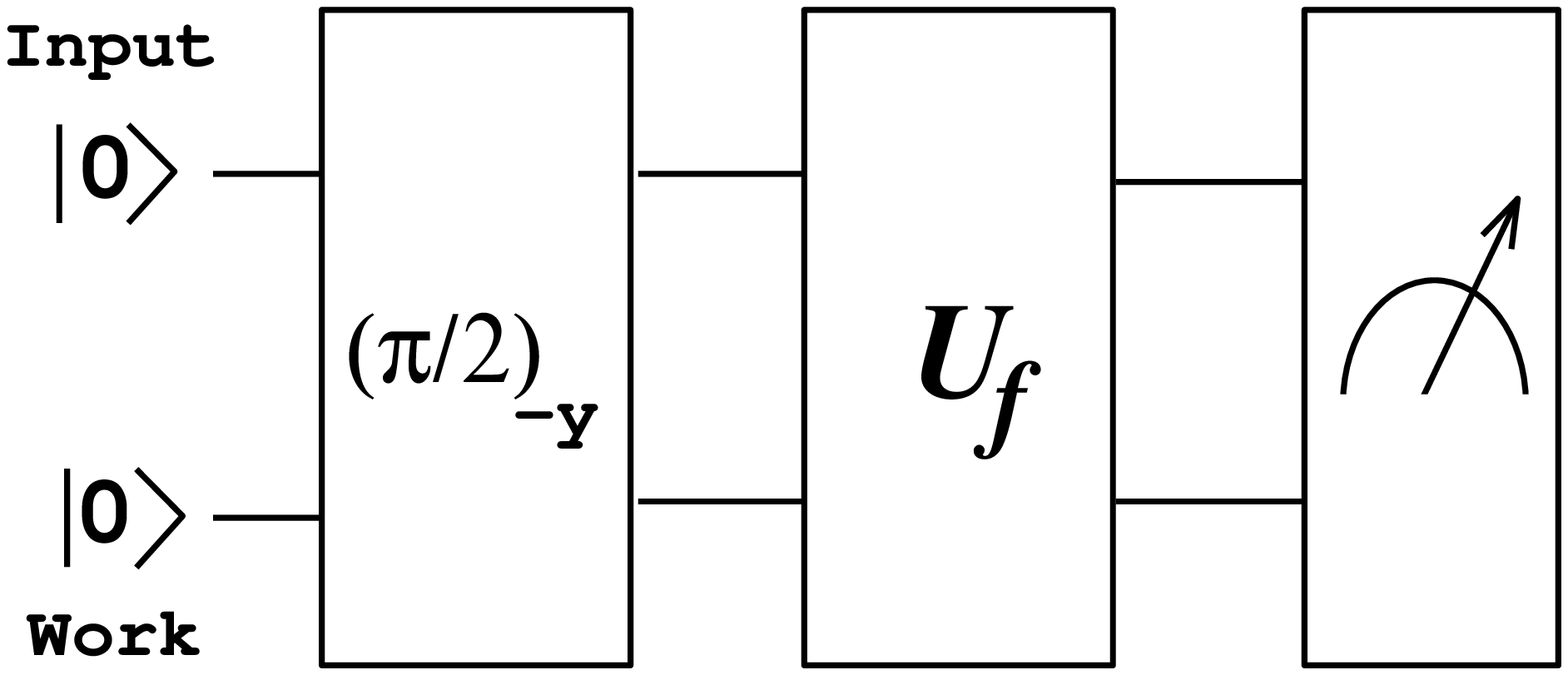,height=16cm,angle=270}
\end{figure}
\vspace{2cm}
\hspace{7cm}
\huge{Figure 2}
\newpage
\begin{figure}
\epsfig{file=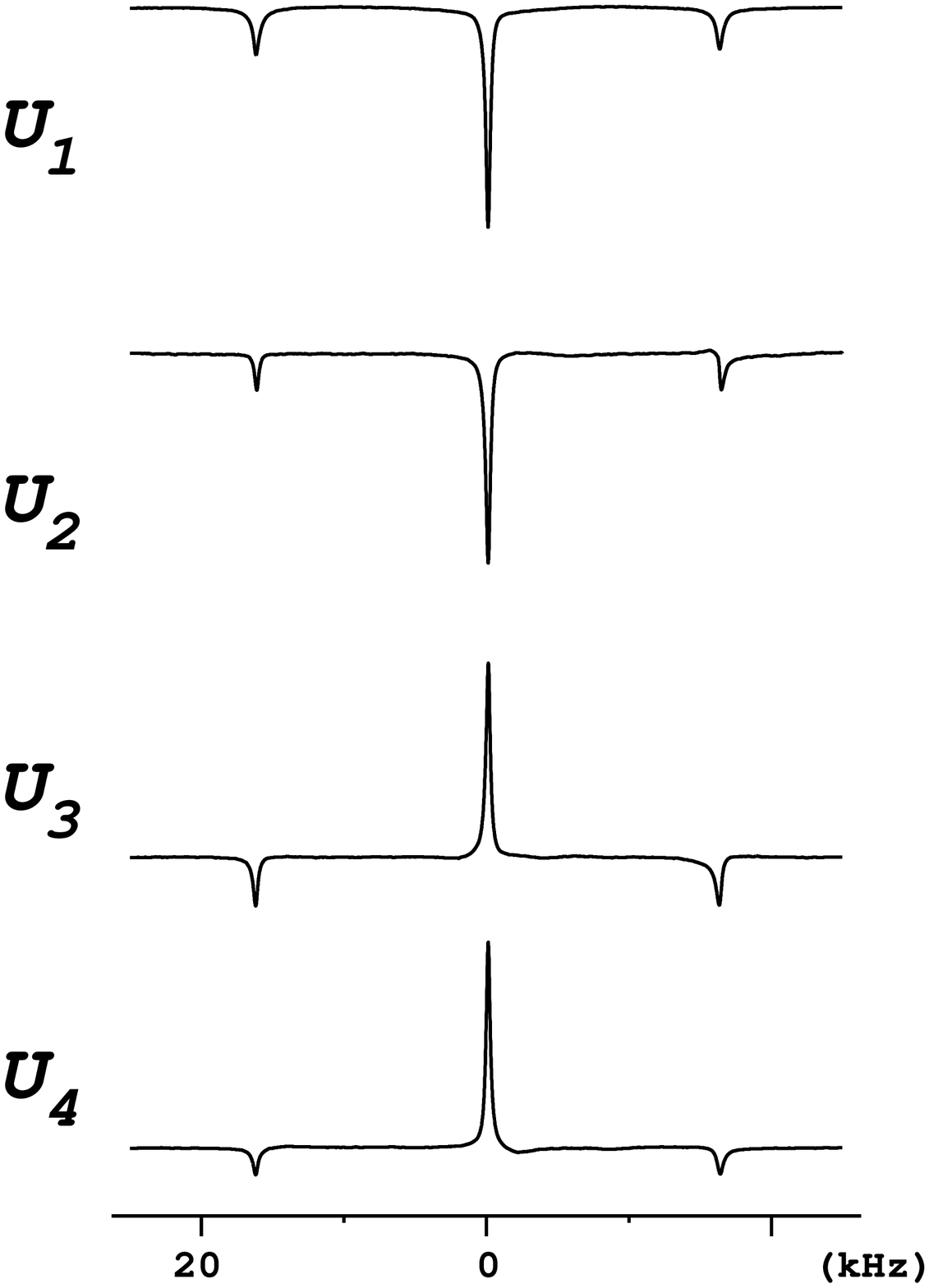,width=10cm}
\end{figure}
\vspace{2cm}
\hspace{4cm}
\huge{Figure 3}
\end{document}